# Workshop on Quantification, Communication, and Interpretation of Uncertainty in Simulation and Data Science

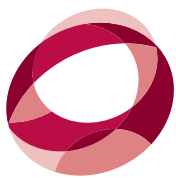

**CCC**

Computing Community Consortium
Catalyst

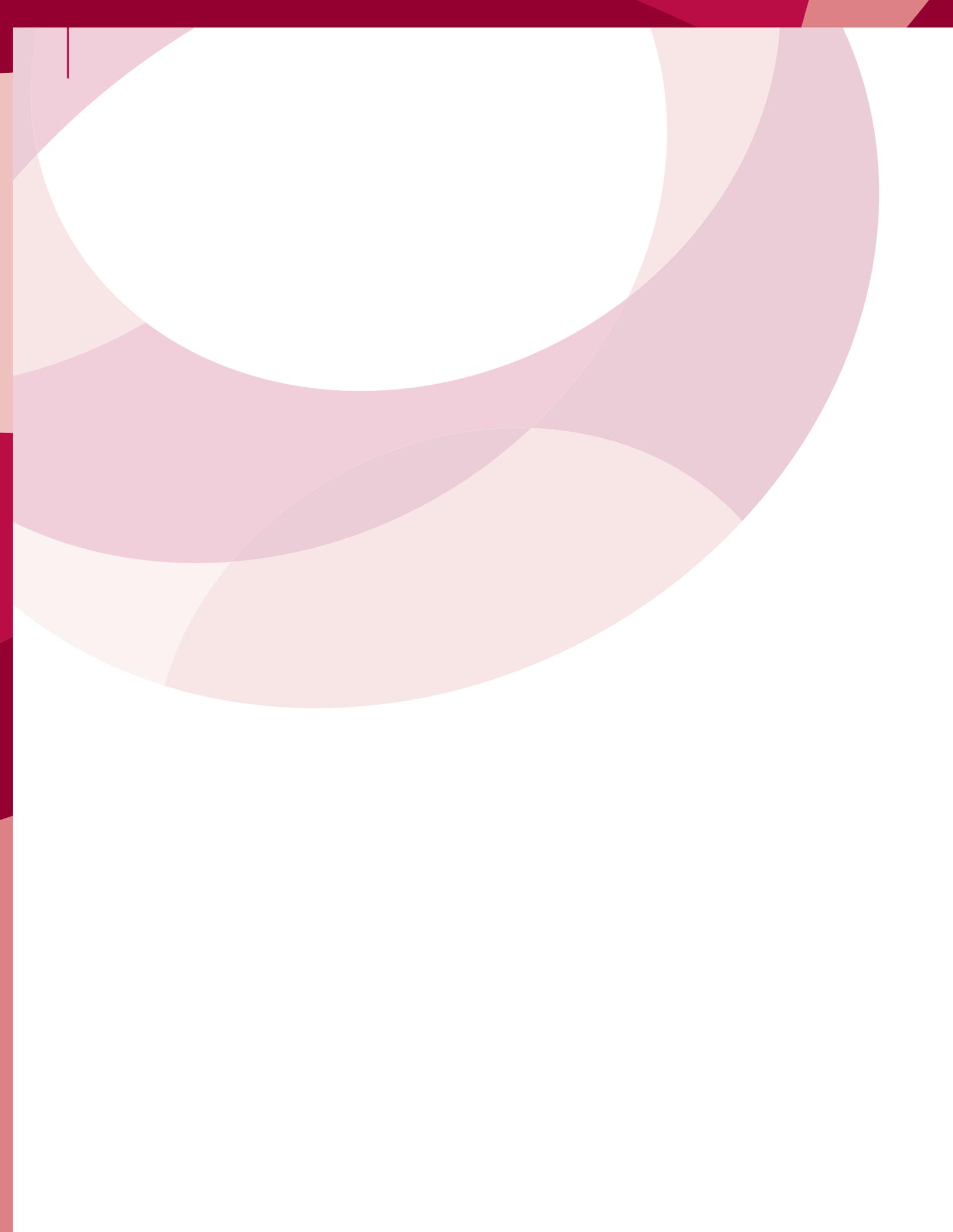

# Workshop on Quantification, Communication, and Interpretation of Uncertainty in Simulation and Data Science

Ross Whitaker, William Thompson, James Berger, Baruch Fischhof, Michael Goodchild, Mary Hegarty, Christopher Jermaine, Kathryn S. McKinley, Alex Pang, Joanne Wendelberger

Sponsored by

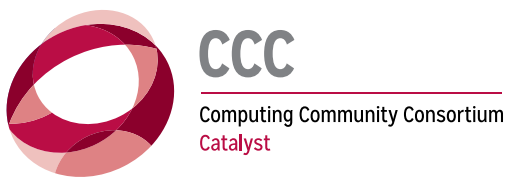

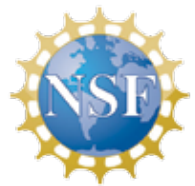

This material is based upon work supported by the National Science Foundation under Grant No. (1136993). Any opinions, findings, and conclusions or recommendations expressed in this material are those of the author(s) and do not necessarily reflect the views of the National Science Foundation.

## Abstract

Modern science, technology, and politics are all permeated by data that comes from people, measurements, or computational processes. While this data is often incomplete, corrupt, or lacking in sufficient accuracy and precision, explicit consideration of uncertainty is rarely part of the computational and decision making pipeline. The CCC Workshop on Quantification, Communication, and Interpretation of Uncertainty in Simulation and Data Science explored this problem, identifying significant shortcomings in the ways we currently process, present, and interpret uncertain data. Specific recommendations on a research agenda for the future were made in four areas: uncertainty quantification in large-scale computational simulations, uncertainty quantification in data science, software support for uncertainty computation, and better integration of uncertainty quantification and communication to stakeholders.

## Workshop Participants

Lorenzo Alvisi, Kate Beard, James Berger, Michael Berry, Derek Bingham, Ronald Boisvert, Valentina Bosetti, David Broniatowski, Eva Chen, Ann Drobnis, Fariba Fahroo, Baruch Fischhoff, Michael Goodchild, Greg Hager, Joe Halpern, Peter Harsha, Mary Hegarty, Vasant Honavar, Donald House, Charles Jackson, Christopher Jermaine, Christopher Johnson, Eugenia Kalnay, George Karniadakis, Tracy Kimbrel, Mike Kirby, Konstantin Krivoruchko, Sandy Landsberg, Steven Lee, Debbie Lockhart, Alan MacEachren, Kathryn S. McKinley, Mahsa Mirzargar, Lace Padilla, Alex Pang, Christopher Re, Tom Russell, Elaine Spiller, William Thompson, Helen Vasaly, Elke Weber, Joanne Wendelberger, Ross Whitaker

# Executive Summary

**Overview:**

Decisions are increasingly made based on information derived from computational simulations or extracted and distilled from large collections of data obtained from diverse sources. Such information is inherently imperfect. Rational decision making is only possible if the level of uncertainty in the source data is both known and incorporated into the decision making process. Despite the importance of dealing effectively with uncertainty, existing approaches have significant shortcomings that will only get worse. The workshop explored these issues and made recommendations for a future research agenda that addresses the need for processing and presenting uncertainty in data.

Three current trends make it imperative that action be taken now. The decisions we must make in multiple areas affecting health, safety, and well being are increasing in complexity and consequences. The dramatic increases in computational resources make far more sophisticated models and inference systems possible, but increasing system complexity makes it harder to quantify and reason about uncertainties in these systems. An explosion of available data has taken place, providing far more information than previously available on which to make decisions, but bringing with it far more ways in which that information can be incomplete, imprecise, or simply in error.

Complicating the development of better methods for dealing with uncertainty is the fragmented nature of the workflow and expertise involved. Lessons learned in simulation science have yet to be applied to data science. Much of the work developing methods for communicating uncertainty is poorly integrated with either uncertainty quantification or the needs and abilities of decision makers or other stakeholders. As a result, success will require a broad based multidisciplinary effort, involving development of a comprehensive set of foundations for representing and communicating uncertainty arising from computational processes that accounts for all aspects of the problem—including the applications, the numerics, the visualizations, the programming languages and computer systems, and the comprehension by users—in a holistic, systematic manner.

**Actionable Recommendations:**

The current state of affairs in the quantification, communication, and interpretation of uncertainty in simulation and data science is creating critical challenges, but it also presents important opportunities. Workshop participants identified four directions in which the research and academic community can have great impact:

◗ There is growing concern that the statistical models currently used to quantify uncertainty in the outputs of simulations won't scale, particularly to large, heterogenous computations models. This leads to a critical need to transition research in uncertainty quantification of computational systems from the analysis of components to the analysis of large-scale systems of interacting components.

◗ The emerging field of data science is largely lacking in generalizable methods for quantifying the uncertainty in the output of analysis systems. As a result, a major new research initiative needs to be initiated in this area. Since data science programs are just getting established in universities, this effort needs to be accompanied by relevant curriculum development.

◗ The increasing use of large-scale computational and data-based analyses in decision support and the increased importance of considering uncertainty in such systems will create substantial burdens for software developers. A major new effort needs to go in to the building of generally applicable, easy-to-use software development tools supporting the representation and analysis of uncertainty.

◗ The fragmented nature of expertise in quantification, communication, and interpretation of uncertainty will become more and more problematic as the scale of problems, the scale of computational resources, and the scale of data continues to increase. It is essential that a major new research initiative be undertaken in communicating uncertainty about large-scale systems to stakeholders in a comprehensive and integrated manner.

# 1 Introduction

In September 2008, the collapse of Lehman Brothers almost brought down the world's financial system [Effect of Financial Crisis 2013]. It took enormous taxpayer-financed bail-outs to shore up the financial industry, and the world economy is still recovering over seven years later. Among the many causes that led to the crisis was a failure among financial regulators, who relied on computational models that did not adequately account for the risks associated with contemporary mortgage and banking practices and consumer behavior.

In March 2011, an earthquake and subsequent tsunami resulted in a nuclear disaster in the Fukushima Nuclear Power Plant, resulting an a meltdown of three of the plant's reactors, the evacuations of more than 300,000 people, the release of nuclear materials into the atmosphere, and long-term contamination of land and water resources. It is estimated that lingering health effects and environmental clean up will last for decades. This catastrophic failure happened despite computational models that predicted the nuclear facility and its surrounding sea walls could withstand a variety of *worst case* scenarios [The Tokyo Electric Power Company].

On the morning of August 29, 2005, Hurricane Katrina struck the Gulf Coast of the United States. When the storm made landfall, it had a Category 3 rating. It brought sustained winds of 100–140 mph and stretched approximately 400 miles across. The storm did a great deal of damage and its aftermath was catastrophic. Levee breaches led to massive flooding and hundreds of thousands of people were displaced from their homes, causing more than $100 billion in damage. In the days prior to the storm, forecasts from the National Hurricane Center, aided by computer simulations of the storm, predicted the magnitude and position of the storm to remarkable accuracy [Katrina Forecasters 2005]. These forecasts anticipated a risk to the integrity of the levees surrounding the city and warned of "incredible" human suffering. By many accounts, these forecasts and warnings were not properly heeded by government officials and the local populace.

The links between engineered systems and computational estimates or forecasts are ubiquitous. In October, 1993, static load tests of the C-17 Globemaster military-transport plane showed that they failed below the required 130% of maximum operating load. This necessitated a redesign, costing tens of millions of dollars and affecting the load and range specifications of that aircraft. The problem was attributed to *optimistic* computational models, which did not properly account for uncertainties or unknowns in the complete system. In early 2012, cracks were found in the wings of several superjumbo, Airbus A380 aircraft, prompting European authorities to order the entire fleet to undergo detailed inspection. The original design of the A380 wing, the potential consequences of the cracks, and the proposed structural repairs [A380 Wing Modifications 2013] and redesigns were all evaluated using computational models.

Meanwhile, computational models are being used for systems at global scales. As scientists, citizens, and policy makers consider the impact of human behavior on climate and the prospects of a warming planet, they rely on sophisticated models of global climate. These climate models entail a system of interacting simulation components including such diverse phenomena as ocean currents, solar flares, and cloud formation. These components and their interactions include dozens of modeling assumptions and parameters that affect the forecasts and their relationships to policy decisions and human behavior. In these circumstances, the efficacy of the computational models and a full understanding of their limitations becomes critical.

The large-scale effects of computer models are not limited to physical simulations. As businesses and governments take advantage of the very large databases, they look for patterns that indicate trends or distinct categories of behaviors in complex agents, such as systems or people. In many cases, unusual patterns are a trigger for further action, and where the stakes are high, as in national security, the consequences of acting or not acting are significant. Even beyond privacy, these data mining paradigms raise concerns among data scientists. The inherent properties of large data sets and the algorithms for finding patterns will invariably lead to the potential for *false discoveries*—cases where large amounts of incomplete or noisy data suggest a phenomenon or behavior which may not be true.

We live in the data age. Modern science, technology, and politics are permeated by information that comes from people, measurements, or computational processes. Thus, most important decisions are made on the basis of data that have been processed and presented by computers. The greater availability of data, along with the tools to acquire, store, and process it, promises more informed, objective decision making. However, data are inevitably imperfect. They are often incomplete, corrupt, or lacking in sufficient accuracy and precision. While consideration of these uncertainties would seem essential to rational decision making, explicit consideration of uncertainty is rarely part of the computational and decision making pipeline. Indeed, the opposite is true—data that is processed and presented on computers often have an implicit connotation of precision and certainty.

It is important to recognize the difference between this emerging world of abundant digital data and the longstanding traditions of science. Most of the great scientists of the past were empiricists, relying on observations and measurements made either by themselves, or by people known to them and trusted. Now, however, the reliability of much of the data on which critical decisions are make is unknown. Data distributed by government statistical agencies such as the Bureau of the Census are well documented, with rigorous procedures of quality control. But the vast flood of data that is now available from groundbased sensors, social media, citizen science, and other unconventional sources is rarely well documented, often lacking in information about provenance, and unlikely to have been sampled according to any recognized sampling scheme. Until effective uncertainty quantification is available for such source data, the investigator wishing to use such data is left with a simple choice: reject the data, or take the risk of trusting them; and is unable to offer much in the way of confidence limits on the results of analysis.

This white paper addresses the compelling need for improved technology to characterize and communicate the inherent uncertainties in complex computational models and large data sets. The timeliness of this need is a consequence of several important, recent trends in technology, policy, and human behavior.

*The increasing scale of problems.*

The first trend in computational modeling is in the scale of human systems and their consequences. Not only do humans build and operate highly complex systems, their effects are widespread. Thus, a single airplane design can transport tens of thousands of passengers each day. Mortgage regulations in a few countries can affect the entire world economy. Nuclear power plants can disperse radiation for thousands of miles, making their reliability a global concern. Perhaps the most compelling example is climate, where policies and human behavior have the potential to dramatically change the world for centuries to come. As the impacts of decision making processes expand, the processes themselves take on more complexity, and the demand increases for computer modeling to help make informed decisions.

*The increasing scale of computational resources.*

Another important development that drives the increase in complex computer models is the rapid expansion of computing resources. Computational power itself has expanded by several orders of magnitude over the last decade, allowing models such as physical simulations to approach a level of fidelity that provides useful, credible input into decisions about complex systems. Additionally, the algorithms and mathematical models that describe the real world have also improved. So, while research and development in modeling continues, for many applications, modeling technology has crossed a threshold of realism and credibility, making it a practical, cost effective way to formulate and analyze important decisions.

*The increasing scale of data.*

The third important development in computer modeling is the availability and the power of data. As large amounts of data become more important in decision making, *data science* is emerging as a distinct discipline, with new technologies, opportunities and academic programs. While a great many approaches have already been developed for data analytics, few involve the systematic incorporation of uncertainty quantification, which has been a growing part of simulation science for the last decade. Thus, there is an opportunity for the field of data science to include uncertainty in its fundamentals, in anticipation of its increasing role in human decision making.

## 2 Decisions Involving Uncertainty in Data Arising from Computational Processes

In planning the workshop and structuring the questions and challenges surrounding uncertainty in simulation and data science, the organizers developed a flow chart of a typical *pipeline* that characterizes the activities associated with computer-aided decision making in the presence of uncertainty. Figure 1 shows a diagram of this pipeline. On the left we see two boxes representing the sources of quantifiable results. For this report, we distinguish the two cases of data coming from computer simulations (*simulation science*) and data harvested from other sources such as large databases (*data science*). In both cases, analysts and modelers examine the output of a model or collection of algorithms applied to data. (Increasingly, we are seeing examples of simulation science operating on information originating from data science methodologies.) The information is then communicated to decision makers, and then possibly further digested and formatted again for the final consumers of the information.

To better understand this process we might consider the particular case of hurricane forecasting, as conducted by the National Hurricane Center. Meteorologists have forecasting models that give predictions of hurricane behavior including its future path, intensity, wind speeds, and storm surge. [National Hurricane Center 2009]. Hurricane forecasts, like many meteorological forecasts, typically consist of a collection of forecasts made by different computational models, sometimes referred to as an *ensemble*.

An ensemble of forecasts helps to characterize the anticipated error or inaccuracy in the set of predictions. These ensembles of forecasts are evaluated by computational meteorologists for consistency, patterns, irregularities, etc. Based on these ensembles and experience with the error in previous forecasts, modelers, meteorologists, and their supervisors meet and decide on the forecasted hurricane track and associated *uncertainty*, which is presented to a wider audience, and eventually the general public, as a *track forecast cone*, as shown in Figure 2. Policy makers use these forecasts to make decisions about the allocation of resources, the implementation of emergency procedures, and informing and advising the public. Ultimately, individuals living in an affected areas must use these forecasts and the associated warnings, orders, etc., to decide on how they will prepare and respond.

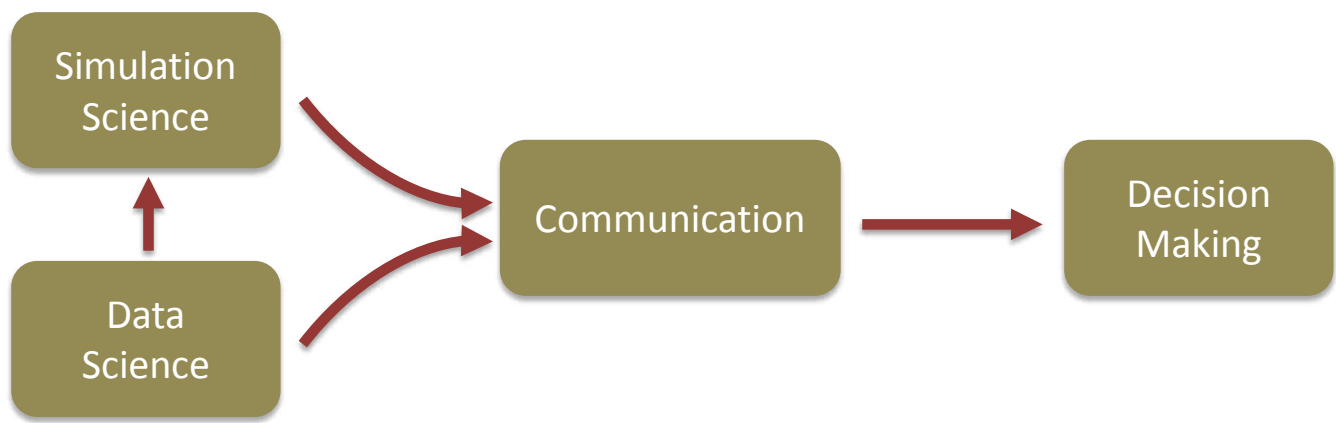

*Figure 1: Data-to-decision pipeline.*

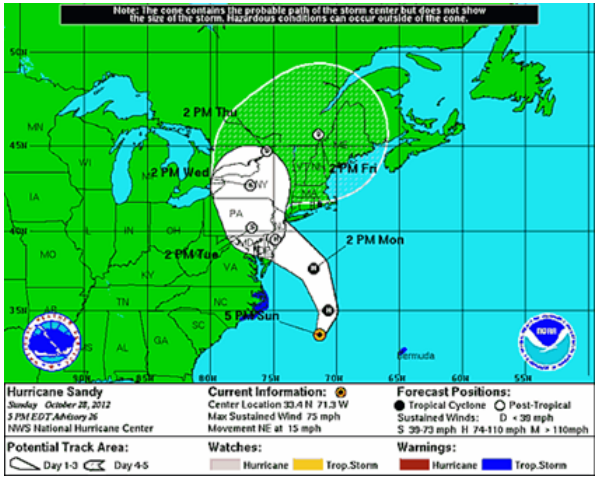

*Figure 2: Track Forecast Cone.*

When considering the *sources* of data, on the left in Figure 1, it is important to understand the nature of both simulation science and data science, which share some common characteristics. For instance, the overall work flow is similar. With both, the data needs to be analyzed, reformulated appropriately for end use, and then communicated to decisions makers. Both cases require analyses and formulations of uncertainty that aide decision makers. Both cases need ways to represent and communicate about the results of large, complex computational processes.

However, there are some important differences between uncertainty in simulation versus data science. Most of the uncertainty in simulations results from combination of uncertainties or errors in the mathematical models, uncertainties in the parameters that underlie those models, or inherent variability in natural phenomena [National Research Council 2012]. The field of *uncertainty quantification* in simulation and numerical analysis is becoming well established as an area of active research, with dedicated journals, etc. The field of *scientific visualization* has actively pursued some of the open problems and challenges associated with producing visual depictions of uncertainty in the often complicated outputs of simulations, such as fields of pressures or velocities, or geometric paths or surfaces.

The field of data science is generally less developed than simulation science in its treatment of uncertainty (although here, too, there is active research in the area, e.g., [Chatfield 1995; Hammer and Villmann 2007;

Bendler et al. 2014]). Uncertainties in the analysis of large data sets arise directly from the data in the form of errors in measurements or inputs, as well as missing, incorrect, or incomplete data. Uncertainties also arise in the modeling of data in machine learning, and result in inherent uncertainties regarding the ability of clustering, correlation, regression, classification, and other procedures to predict or model unseen data. Some of this uncertainty reflects inherent stochasticity (as in simulation-based forecasting) in processes being modeled or predicted. For instance, when studying human behavior, either groups or individuals, one would expect limited accuracy because of the inherent complexity of humans and incomplete observations. In addition to inherent stochasticity, the application of modeling or machine learning algorithms to large datasets faces important (and exciting) unresolved issues in the use of cross validation and testing data, with resulting biases and over fitting that limit their accuracy on new datasets. Likewise, data modeling algorithms include parameter or *model* choices that affect the results, and therefore need to somehow be accounted for when these results are used to make decisions.

Of particular importance in data science is the effect of false positives in analyses of large datasets. Many data science tasks entail the *detection* of particular, often unusual events through indirect measurements [Chandola et al. 2009]. This would be the case, for instance, in certain security applications, where the intentions of an individual or group are inferred indirectly through patterns of behavior. These behaviors might entail certain kinds of communication, purchases, or patterns of travel. However, as data sets and number of analyses grow larger, the probability of *false detection* increases, thereby leading to uncertainty about whether the result of a detection algorithm is actionable, given the decision maker's objectives. The ability to capture and communicate information about false detections will be an important problem as the applications of machine learning to large data sets mature and become more widely used.

Uncertainty analysis in data science and simulation must ultimately serve decision maker's needs. Effective decision making requires considering uncertainty both *within* and *between* each step, and then communication of relevant information about the reliability of the outputs of these processes to decision makers in authoritative, comprehensible, and actionable form. Uncertainty quantification for computational simulations is a maturing discipline, but little study has yet gone into the communication of its results.

Data analytics is rapidly becoming far more sophisticated and enjoying widespread use, but is still largely lacking in well principled methods for quantifying and communicating the uncertainty associated with the information contained in large data sets. While communicating uncertainty to decision makers has been studied in the geospatial, visualization, and cognition communities, effective and generalizable methods are still largely lacking. The field of decision science has extensively studied decision making under uncertainty [Fischhoff and Kadvany 2011; Howard and Matheson 2005; Kahneman 2011; Morgan and Henrion 1990; National Research Council; O'Hagan et al. 2006], but this work has yet to be integrated with either formal uncertainty quantification or the explosion of computational uncertainty associated with data analytics.

One challenge to addressing uncertainty in data science is the relatively new state of the field. That challenge also creates an opportunity to shape an emerging field so as to accommodate the relevant disciplines from the start. Educational programs in *data science* are just now popping up in universities. The increasing demand for people trained in data science suggests that their number will increase. Thus associated fields such as computer science, statistics, and decision science are in a position to shape their curricula and influence their pedagogy to include training in analytical and behavioral aspects of error and uncertainty.

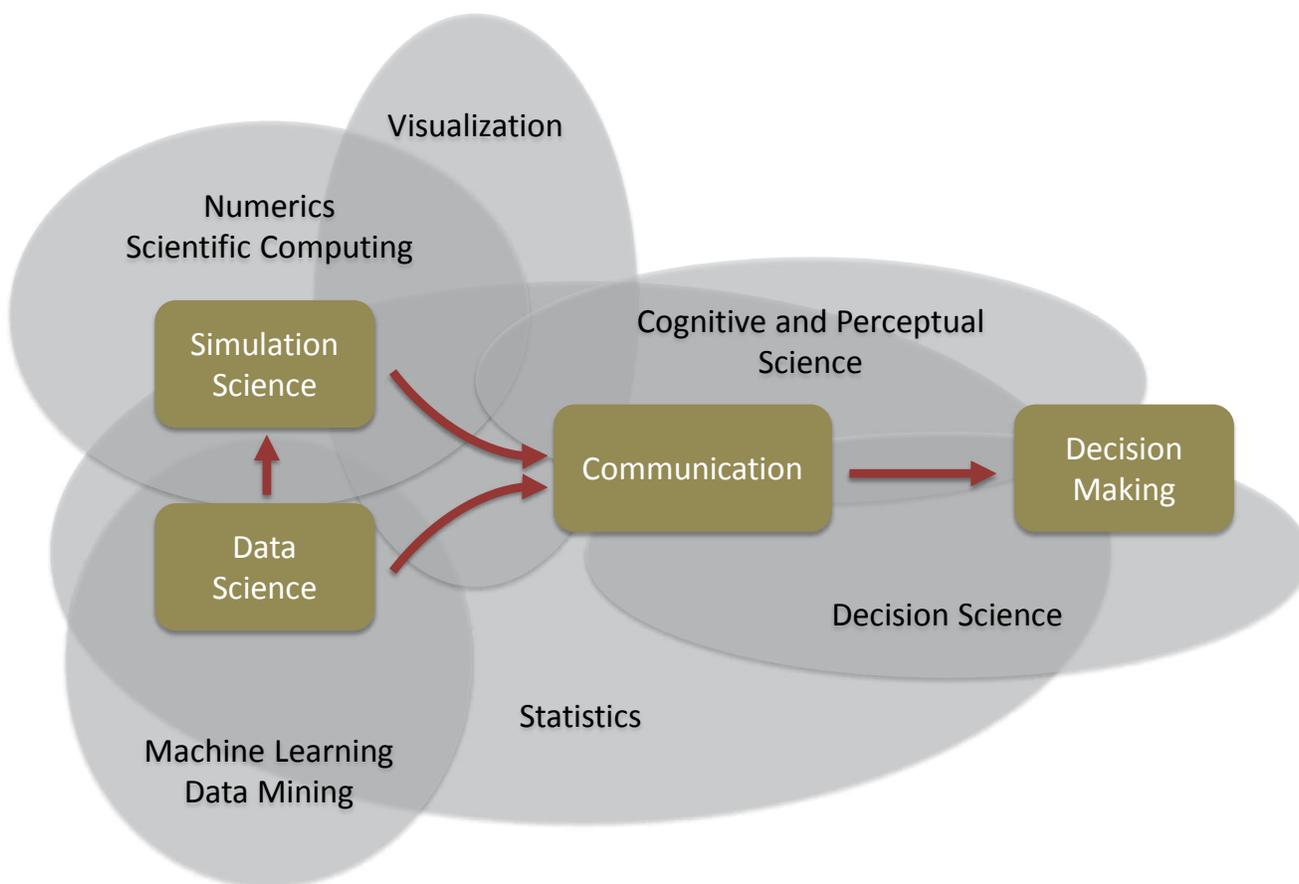

*Figure 3: Academic disciplines relevant to understanding the data-to-decision pipeline.*

Figure 3 highlights an important issue in addressing technical challenges associated with uncertainty in computation, which is that the relevant research is fragmented and does not map well onto the individual steps in the data-to-decision pipeline. These current interactions of different fields with the work flow are depicted as grey ovals in the figure.

For instance, simulation science deals mostly with large-scale computation models, and is largely within the purview of numerical analysis, scientific or high-performance computing, and application-specific engineering. Data science, with its emphasis on the analysis of existing data, draws on expertise in AI, machine learning, and data mining. While at a high level, simulation science and data science play similar roles in the overall data-to-decision pipeline, the associated disciplines share little in problem framing, computational tools, or mechanisms for communicating results. Of course, *uncertainty* is fundamental to the field of statistics, and parts of the statistics community have been actively involved in the quantification of uncertainty in both the simulation and data contexts. However, tighter coupling is warranted in addressing both fundamental problems and technologies for quantifying uncertainty in data and simulation science.

The communication of uncertainty to decision makers and the manner in which they utilize that information is studied by a disparate set of academic disciplines. Scientific visualization focuses mostly on pictorial representations of uncertainty associated with the high-dimensional outputs of physical simulations, such as pressure fields or velocity fields [Pang et al. 1997; Potter et al. 2012]. The information visualization community has dealt with the issue somewhat less, reflecting the similar situation in data science. Although cognitive and decision sciences have vast basic research literatures and substantial applications in many domains (e.g., health, environment, finance), they have had relatively little connection with computer science—outside the important work on human-computer interaction.

## The Difficulty in Interpreting Statistical Quantities

It is clear that our understanding of uncertainty is impacted by the statistical tools that are readily available for analysis and the difficulty that nonexperts have in understanding their meaning. The classic example of this is the p-value. For example: in a recent talk about the drug discovery process, the following numbers were given in a hypothetical illustration. A total of 10,000 compounds of interest are screened for biological activity. Of these, 500 pass the initial screen and are studied in vitro. Twenty five pass this screening and are studied in Phase I animal trials. One compound passes this screening and is studied in a Phase II human trial. This scenario is consistent with moderately noisy testing schemes and a set of entirely *inactive compounds*, with screening based on significance at the 0.05 level. With increasingly large amounts of data being mined, this *multiplicity problem* can easily lead to "discoveries" that are in fact just artifacts of noise.

Conveying and understanding the power and limitations of statistical analyses is no small task, and it becomes more difficult as we consider multiple related experiments on a given set of hypothesis or multidimensional outcomes. On the other hand, there is evidence that visualization helps convey the uncertainty in these outcomes. Meanwhile, there is a growing sense of a need for alternative methods for quantifying and communicating such results.

The understanding of uncertainty in quantitative data and its use in decision making is an area of study in perceptual and cognitive psychology as well as decision science. While psychologists study the human capacity for reasoning under uncertainty, they rarely consider the kinds of complexity found in modern simulations and data analytics. The field of decision science takes a more integrated approach involving formal analysis of systems, empirical studies of human behavior, and interventions for improving design. However, ties with the engineering, machine learning, and visualization communities are lacking. Creating these connections will raise fundamental questions for all fields involved. Hence, there is a growing need for behavioral studies of how people deal with complex data types and alternative representations of uncertainty. Creating these connections will raise fundamental questions for all fields involved and will result in the need for fundamentally new methods for quantifying uncertainty in a way that dovetails with communication, understanding, and decision making. The current fragmented nature of the relevant disciplines makes clear the need for an intrinsically multidisciplinary approach that considers uncertainty in computation in a holistic, end-to-end manner.

## 3 Challenges and Opportunities

In this section we describe the challenges, technical and otherwise, facing improvements in the use of uncertainty in computation, as well as opportunities that will result from solutions to those challenges and the efficient and effective use of uncertainty in computer-aided decision making.

### 3.1 Simulation Science—Complexity, Scale, and Verification

While the field of uncertainty quantification in simulation science is somewhat advanced, effective, widespread use is limited by several factors. One important factor is complexity. Large systems that are subject to simulation-based design and testing, such as airplanes or power plants, are extremely complex systems with thousands of interacting components. The state-of-the-art uncertainty quantification technologies deal primarily with a single, physics simulation, such as the mechanical or thermal behaviors of a part or a relatively small collection of parts in proximity. The interactions of thousands of parts, connected by physical proximity, electrical connections (wireless or wired), energy conduits, or a dependence on a common physical resource (e.g., air, water), are still well beyond the state of the art in uncertainty quantification. The importance of these interactions became especially critical in the case of the Fukushima disaster, where the failure of separate primary and backup energy systems (from the same cause but different mechanisms) interacted to facilitate the core failure.

Several challenges are important in dealing with complex, multicomponent systems. First is the computational scale, and the limitations this places on the ability for simulations to span the appropriate set of possible outcomes of these systems. One strategy is to encapsulate systems, characterize their error or uncertainty separately, and then model their interactions and the associated propagation of uncertainty. This strategy entails thousands of interactions and significant heterogeneity in the types of interactions. These interactions are not merely additive, and they will demand new numerical, statistical, and computational tools for representation and computation. These interacting subsystems inevitably include feedbacks and nonlinearities that can lead to emergent behaviors that are not easily modeled by the same approximations for each system. Thus, this massive and complex data may require new statistical tools, beyond the capabilities of current well researched tools.

Another important aspect of complex systems is verification. Verification entails the evaluation of computational models against empirical data. For isolated, physical systems, such verification would typically entail laboratory experiments—for example, to verify that the difference between mechanical behavior of a part and its computational model are within predicted uncertainty. More complex systems do not easily lend themselves to controlled, laboratory experiments, and therefore other verification methods must be pursued. For instance, in fields such as meteorology and finance, *hindcasting* is used to validate computational models against historical data. Thus, given initial and boundary conditions, hindcasting validates new prediction mechanisms with prior events, for instance, validating weather forecasting models with data from the summer of 1984. This strategy becomes more challenging when considering design of artifacts, systems, or policies, or in cases where recorded history does not readily provide a sufficiently general set of examples or sufficient data. This concern has been raised in the issue of climate modeling, for example, where comparisons against several decades of climate data are used to verify a phenomenon which varies, arguably, on much larger times scales. This use of historical data is also a concern in economics, where both rare events and changing operational regimes (e.g., political or cultural changes) challenge the ability of models to generalize into the future. While progress has been made, the technical challenges for verifying large, complex either man-made or natural systems remains an important problem.

The complexity of the systems and the reliance on historical data for modeling and verification raise other important issues in the quantification of uncertainty. Many statistical models include characterizations of outcomes that entail typical or representative outcomes with an associated notion of variability or deviation from this typical outcome. This philosophy is encoded in the typical formulation of a *forecast* (e.g., from a simulation) and an *error*, which assumes some deviation from the forecast with a decrease in probability as potential outcomes deviate dramatically from the forecast. Thus, events that are extremely different from the forecast are considered to have negligible probability, and are often considered inconsequential. However, these, so-called *rare events* are often the events that pertain to catastrophic outcomes when considering decisions about policy or design—thus, they are the most interesting cases to evaluate. Closely related to this phenomenon is the idea of a *tipping point* in the behavior of a complex, nonlinear system, where events (usually rare) force components of a system outside of their normal operating ranges, resulting in very unusual (often undesirable) behaviors. This observation suggests a need for new paradigms in modeling behaviors of systems to account for rare events. For instance, instead of discounting rare events as not probable, one might instead study the situations (e.g., external events or parameters) that would cause a system to fail and then quantify, in a systematic manner, just how possible such events might be—using not only historical data but first principles of the physical phenomenon that govern the inputs to these systems. Thus, in the case of Fukushima, one might ask what size of tsunami would result in catastrophic failure, and then simultaneously study the nature of that failure and whether such a wave is possible, in principle, given the oceanographic and tectonic context of that region. The development of such new paradigms, as well as understanding their theory and practical instantiations, is critical for the successful use of simulation science for large-scale systems and decision making.

Dealing with the challenges above entails computation at a scale that grows with both the complexity of the systems and the widening range of events that one is considering. Thus, one can imagine not only large scale simulations, but a large number of simulations in order to discover situations that produce nonlinear behaviors, tipping points, and catastrophic behaviors, with enough frequency to be able to characterize their behavior. Thus, there is a growing need for computational resources and software infrastructures that allow data scientists and statisticians to readily acquire the data they need. While this challenge will continue to grow as simulations become more sophisticated, there is an opportunity to establish the infrastructure for studying these issues with state-of-the-art sized problems, so that these new capabilities in simulation science will grow as the field develops.

### 3.2 Uncertainty in Data Science

The field of data science is, of course, vast. Here we consider the recently developing trends toward analysis of large data sets with an eye towards:

▶ detecting events, signatures, and patterns (based on training data or based on predefined patterns or anomalies),

discovering the underlying relationships between data points or parameters (e.g., clustering, correlation, dimensionality reduction),

transforming data (processing data into another form, such as summary statistics), or making queries (often approximate) within very large databases.

The class of tasks, types of data, and challenges in dealing with uncertainty in data science are heterogeneous and complex. Here we try to put some structure on the challenges and opportunities, with the understanding that the field is in its very early days.

The uncertainty in the estimates or findings derived from large-scale data analyses typically derive from several factors. The first contributing factor is the source data itself, which is subject to typical measurement noise (e.g., the physical location of an entity or sensor fidelity), but is also incomplete and suffers from certain kinds of *replacement* noise, where some data points are simply "wrong" (e.g., some data is corrupted). Furthermore, much of the data which is considered *source data*, is actually derived data, from preprocessing, such as the extraction of keywords or semantics from text documents. The algorithms that perform these transformations are not entirely robust, and thereby introduce their own kinds of noise into large databases. Certain kinds of queries on very large, possibly distributed, databases, will be tractable only if they are processed in an approximate manner. Queries such as aggregate statistics on subsets of data or proximity queries, such as nearest-neighbor lookups, will be most efficiently answered in an approximate fashion with error bounds that typically decrease with execution time. Coherent, general, statistical models of these various phenomena, which result in uncertainty in data bases (and data queries) remains an open problem. The consensus is that such models or methods are essential to propagating these uncertainties through various analyses that result in quantitative outputs.

The typical data analytics pipeline processes databases to derive quantitative results in one of several different ways. One pipeline relies on classifiers that are trained and tested on specific datasets, often before or during *deployment in the wild.* Classifiers are essentially regressors on noisy data with nondeterministic outcomes, and therefore have expected rates of error—which can, in principle be estimated from test data. However, there are fundamental problems with the construction of realistic training and testing data, and these data sets can introduce biases that may result in optimistic estimates of error. A significant amount of research has addressed the problem of biases in training data and online training to address shifts in data sets over time [III and Marcu 2011]. However, while machine learning algorithms are maturing and computational resources are improving tractability, data sets are expanding in size, scope, and application. As each progresses, the inherent limitations of the data, the training methods, and the uncertainty in the outputs will be a critical component of practical uses of these methods in important applications.

Another class of data analysis pipeline entails the use of *unsupervised* algorithms, that infer structure in datasets or relationships between data objects, without explicit examples of correct or incorrect results. In these kinds of analyses, data integrity and inherent stochasticity of the underlying phenomena continue to play a role. However, the absence of training data exposes a greater sensitivity to modeling choices. For instance, the result of a clustering or dimensionality reduction is not a stand-alone *answer* to a question, but depends on a set of choices on the types of models and parameters within these models that one uses. As people interpret outputs from an unsupervised algorithm, it will be important to understand the extent to which these are features of the data or results of modeling choices. Some work has been done in the area of clustering, for instance in analysis of ensembles of clusters [Topchy et al. 2005], but this very important work is quite early, and there is neither a general methodology that applies across different types of algorithms or systematic methods of summarizing and presenting these types of uncertainties.

Data analysis pipelines are often used to discover relationships between data, such as group differences

(e.g., the p-value example in Section 2), correlations, or anomalies. Such *discoveries* are a critical part of the trend toward *data-driven science*, but they present the field with some important technical challenges. As the p-value example presented earlier demonstrates, typical methods for examining the significance of hypotheses are not well suited to very exploratory analyses. While correction factors for multiple comparisons may be well developed, they often do not scale in a satisfactory manner to thousands or millions of "hypotheses." Likewise, the detection of irregularities or anomalies in very large (and noisy) data sets is prone to false positives, and the uncertainty of the results must be carefully weighed in many of the very serious scenarios (such security and public health) where these methods are beginning to see more widespread use. The challenge is a set of new paradigms or methodologies that change the way we think about and quantify our uncertainties about these types of detections. Decision makers have an opportunity to prioritize decisions, resources, and actions; to learn from their experiences to best utilize resources; to preserve the privacy and integrity of innocent individuals and groups; and at the same time, to detect important opportunities and threats.

The development of large-scale data science will depend not only on the development of new methodologies and computational methods, but also on the availability of data against which to test and validate these methods. Much of the current work is based on either very small or homogeneous datasets or limited access to restricted-use private or protected databases. The lack of availability of sufficiently large, diverse, data sets is an impediment to progress. The challenges are that most *real* data sets have concerns regarding inherent commercial value and/or privacy, while synthetic data sets often lack the desired, size, noise properties, and heterogeneity needed to explore uncertainty. The opportunity in developing such data sets is the development and characterization of entirely new paradigms for quantitative analysis the advances that will help the field of data science to develop a mature set of robust, reliable tools to aid in practical, every-day decision making.

The relatively young status of data science presents another important opportunity with respect to research in uncertainty. Academic institutions are beginning to respond to the growing demand for *data scientists* with associated curricula and/or academic programs and degrees. Universities are hiring new faculty with expertise related to data science, while textbooks and publication venues proliferate. The foundations of the *field* of data science are just now being established, and these foundations are likely to become codified and more stable in the next three to five years. Thus, this represents an important time for the research community to influence and shape this burgeoning field, and establish the quantification and communication of *uncertainty* as a key component of thinking that goes into the design of any analytics system. Systemically incorporating uncertainty will require research investment, sponsorship of venues and tutorials, and the promotion of academic programs that include uncertainty as first-class element of data analysis together with a more holistic understanding of the role of data analysis in computer-aided decision making.

In addition to the computer-aided, decision making pipeline described in Section 2, there are, of course, a wide variety of ways in which computers help in making decisions, at both the large scale and the small. For instance, there are already computer systems that automatically act on uncertain data, for instance, in search and navigation applications, speech recognition, and anomaly identification. These kinds of fine-grained, low-level data analysis tasks are ubiquitous. Meanwhile the programming and software systems used to build these applications (i.e., programming languages, compilers, run time code, etc.) generally ignore uncertainty, making it very challenging for data scientists to develop applications that make systematic use of uncertainty. Programming systems ignore uncertainty because they are typically unable to efficiently represent or reason about this extra information, and therefore provide little help to data scientists in expressing or computing with uncertainty in their data or models.

While researchers have developed some foundations for computing with uncertain values in programming languages [Giry 1982; Ramsey and Pfeffer 2002; Vajda 2014], the theory is incomplete and popular programming languages have not yet adopted these strategies. Currently, the burden on programmers is very high, requiring substantial machine learning and/or statistical expertise as well as significant development effort to program these complex systems correctly. Some recent, domain specific approaches to address this problem for databases [Dalvi and Suciu 2007] and artificial intelligence [Wingate et al 2011] show promise. Meanwhile more general solutions have been proposed that treat uncertain data values as first-order types [Bornholt et al. 2014]. However, substantial additional programming systems research is needed to deliver languages, automated systems, and tools that will help data scientists to develop correct applications that consume and characterize uncertain data, create uncertain models, compose models, and act on the results.

### 3.3 New Computing Paradigms and Uncertainty

As the size and scope of the computational problems we take on grows, uncertainties are introduced in the underlying models, assumptions, and data. Meanwhile, several important trends in computing architectures and scalable algorithms promise to compound the challenges in dealing with uncertainties in the computations that aide decision making. These trends in computing take several different forms, but all lead to the same conclusion: *important new computational paradigms will introduce uncertainty or error into otherwise deterministic computations or algorithms.*

A important recent trend in computer architecture design is *approximate computing* [Han and Orshansky 2013]. The strategy of approximate computing is generally to perform lower-precision computations using either a subset of available hardware (e.g., reduced number of bits in an arithmetic operation) or fewer cycles. Approximate computing designs are proposed primarily for computer architectures that face power constraints, but the approach can address compute-time constraints as well. The strategy applies to both low-level operations, as well as approximate solutions to higher-level algorithms. The *approximate* nature of the results of these computations introduces an error or uncertainty that accumulates, affecting resulting computation.

*Stochastic computing* is a related computing paradigm, which was first proposed in the 1960s but is now seeing renewed interest. Stochastic computing takes advantage of the representation of numbers as random bit streams in a manner allowing simple, fast, and approximate algorithms for arithmetic operations. The stochastic nature of the strategy produces an expected error for each computation, which decreases as one considers longer random strings (more computation). Uncertainty in stochastic computations compound as low-level operations are combined in higher-level, algorithmic tasks.

Related to these approximate or stochastic computing technologies are a range of approximate algorithms for large-scale computational problems that are otherwise deterministic. One example is approximate queries on large and/or distributed databases, which entails using sampling strategies to produce approximate, aggregated outputs of database searches or queries. Such algorithms often come with theoretical or empirical estimates of uncertainty.

While the CCC visioning workshop did not focus on these very specialized computing paradigms, they represent important trends that compound the inherent uncertainties associated with simulations and data analytics at large scales. We anticipate that these factors will contribute to the degree and complexity of uncertainties and uncertainty quantification and communication associated with advanced decision making tools. An important aspect of these uncertain computational paradigms is that *otherwise deterministic computations may have associated uncertainties or errors.* This suggests a need for software support for uncertainty computation. For instance, it is likely that there will be a need for programming languages (or extensions) that support estimating, propagating, and reasoning about the errors or uncertainties that arise in basic, low-level operations, so that software engineers can instrument algorithms for uncertainty with relative ease (or perhaps, automatically).

As in all of the uncertainty computations described in this paper, uncertainty introduced by new, efficient, approximate computing paradigms will need to be properly accounted for within software/algorithms and communicated effectively to decision makers. In this sense, the challenges associated with uncertainty in large computational models align with those introduced by new, approximate computing paradigms. However, these new paradigms introduce an additional, different perspective into the uncertainty process, which is the need to support uncertainty computation in a wider variety of problems and with greater (programmer) ease and efficiency at the system software, programming language, and application layers.

### 3.4 Communication and decision making

In an article in *Science*, Spiegelhalter et al. [2011] lament the absence of research in presenting uncertainty and conclude: "Given the importance of the public understanding of health, economic, and environmental risk, it may appear remarkable that so little firm guidance can be given about how best to communicate uncertainty." The consequences of failing to account for uncertainty are significant. Decision makers often face risks and tradeoffs, with significant ramifications for financial assets, economic development, public safety and health, or national security. These risks must be weighed against the integrity of the model in a way that allows decision makers to include appropriate margins that reflect their priorities. In other cases, the results of these decisions must be conveyed to a broader class of *stake holders*, who may not have access to the computational models themselves, but are often prone to ask "How certain is your prediction?" Furthermore, the credibility of the process itself—of using forecasts based on computational models to make decisions—depends on proper evaluation of models and outcomes *in relation to expected levels of error.* If the expectations of policy makers and stake holders are not calibrated to appropriate levels of uncertainty in computational models, the effective use of these models and their results will be jeopardized. This dilemma has recently made international news in the context of climate science, and a panel examining the work of the Climate Research Unit concluded [Lord Oxburgh 2010]: "Recent public discussion of climate change and summaries and popularizations of the work of CRU and others often contain oversimplifications that omit serious discussion of uncertainties emphasized by the original authors."

Although there is much potentially relevant research in the behavioral and decision sciences, it has yet to be applied and extended to create a systematic approach to communicating the intrinsic uncertainty in the outputs of a simulation or other complex data to users making high-stakes decisions based on such data. Current practice either ignores uncertainty, or in a few cases uses ad hoc techniques to present uncertain data. Considering this problem narrowly, as simply the development of a set of description or illustration methods, is bound to fail. This is because understanding and using uncertainty is notoriously difficult for both trained users and novices [Belia et al. 2005; Tversky and Kahneman 1974]. Furthermore, solutions must apply to a wide variety of situations in which uncertainty plays a role and must account for large differences in user training and skills. A successful approach to communicating uncertainty must create tools and procedures that span the process from initial modeling and quantification of uncertainty to end-user decision making, using systematically validated and methodologically sound procedures to determine their effectiveness.

Psychological studies of decision making focus on the cognitive and affective processes underlying choices under risk or uncertainty, including intuitions and heuristics, computational processes of weighing risks and rewards, and combinations of these processes [Loewenstein et al. 2001; Reyna 2004; Todd and Gigerenzer 2007]. In many studies of decision making, people are asked to reason about simple decision scenarios stated in in verbal or numerical form, with minimal if any information provided about the source of the information, including uncertainty in the data. Other studies are concerned with the communication of scientific information about health risks, natural disasters, and global challenges such as climate change to consumers. These challenges have prompted decision scientists to question how to best communicate scientific data to the general public [de Bruin and Bostrom 2013; Fischhoff and Scheufele 2013; Weber and

**Verbal:**
– There is a fair chance of a market upturn by the election

**Numerical:**
– There is a 20% chance of rain tomorrow; p < .05

**Graphical:**
– Static
– Dynamic
– Interactive

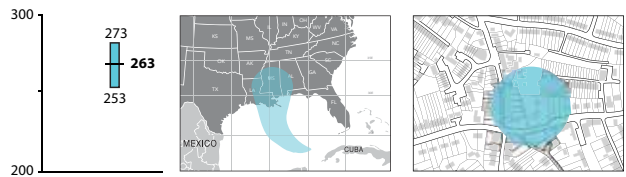

**Combinations** of these (multimedia & multimodal displays)

*Figure 4: Ways of communicating uncertainty.*

Stern 2011]. Current approaches emphasize the need to match scientific communications to people's existing mental models, and to the abilities, and skills of the consumer. However they focus on communicating the results of scientific investigations, rather than the scientific processes that lead to these results, and there is currently much controversy about whether to even present information about uncertainty of the results (e.g., [Joslyn and LeClerc 2013]).

The current failure to systematically account for uncertainty in interpreting complex data stems, in large part, from a lack of knowledge, strategies, and tools for effectively depicting uncertainty. Information about uncertainty can be presented as text, numerical expressions, static images, animations, interactive visualizations, and more (Figure 4). We know that the way in which uncertainty is depicted affects the way in which it is understood [Correll and Gleicher 2014; de Bruin et al. 2013; Finger and Bisantz 2002; Garcia-Retamero and Cokely 2013; 2013; Stone et al. 1997]. The literature is burgeoning on topics such as the effectiveness of alternative displays; the role of trust in communication; and individual differences in decision-making competency [Fischhoff and Kadvany 2011; Fischhoff and Scheufele 2013; 2014]. Although much of that research has been prompted by concern over complex, uncertain decisions (e.g., medicine, technology), it has had little contact with the simulation and data science communities. As a result, there is an opportunity to leverage the progress in these fields.

We also know that the nature of the people performing those tasks have a major impact on which forms of communication are most effective [Galesic et al. 2009; Okan et al. 2012; Reyna et al. 2009]. It is also likely that effective communication of uncertainty will depend on the nature of the tasks that need to be performed with uncertain information. Categories of tasks can range from the need for a general understanding of a data set to the prediction of a specific value. The need for persuasion creates a different set of requirements for effective communication than does the need for decision making. The temporal urgency of the situation also matters (e.g., wildfire vs. ebola vs. climate change). We are still far from having useful, prescriptive theories of any of these effects.

A significant body of literature on visually communicating information about uncertainty has come out of the geospatial and visualization communities [Bonneau et al. 2014; MacEachren et al. 2005]. One key problem in the presentation of geospatial information is that most users expect maps to be perfect, and so resist even the concept of uncertainty. A number of key technical challenges are outstanding as well. There is typically positive spatial autocorrelation in errors, complicating modeling, calibration, and the specification of metadata. Uncertainty causes significant difficulties for data provenance, interoperability, and the choice of basic representational frameworks (e.g., raster vs. vector). Open problems relevant to scientific and information visualization include how to communicate uncertainty in relational data, the potential benefits of using ensemble representations as a communication tool, how to deal with uncertainty in high-dimensional data, and how to communicate information about error propagation.

Complicating the situation is the fact that relatively few comprehensive studies have been done on how well various techniques for presenting information about uncertainty affect final judgments. Instead, the literature is replete with ad hoc *user studies* with no systematic choice of tasks, scenarios and problem framing, participants, or response measures [Kinkeldey et al. 2014]. These user studies rarely rest on sound scientific principles or careful evaluation. In addition, the populations on which they are performed are rarely representative of the stakeholders associated with high-stakes, real-world problems involving substantial uncertainty.

Critical to the problem is the fact that judgments about the effectiveness of alternative approaches to modeling and communicating uncertainty cannot be based purely on the intuitions of users, domain experts or visualization experts. People's intuitions about what makes a good display or information do not always conform to what is objectively a good display for their task [Hegarty 2011; Hegarty et al. 2012; Smallman and St. John 2005], so that designers of user interfaces speak of a *performance-preference dissociation* [Andre and Wickens 1995; Bailey 1993]. For example, users often prefer more complex displays but in fact are better served by simpler displays [Smallman and St. John 2005]. Clearly, more attention needs to be paid to validation of how uncertainty is conveyed to stakeholders, beyond the validation of uncertainty quantification methods themselves.

The field of decision science has made important strides in understanding and facilitating human decision making in real-word settings (e.g., [Fischhoff et al. 2012; Fischhoff and Davis 2014]). The research addresses three issues central to making best use of analytical methods. One is communicating with decision makers, so that analyses are as relevant as possible to their needs and so that their results (and attendant uncertainties) are properly understood, with respect to their decision making implications. The second is translating behavioral research into analytical terms, so that models make realistic assumptions about human behavior affecting system performance (e.g., how people respond to evacuation or quarantine notices, how vulnerable individuals are to phishing attempts, how consistently operators maintain equipment). The third is assessing and improving the human element of modeling (e.g., how teams are constituted, how expert judgments are elicited, which sensitivity analyses are performed, how scenarios are chosen and models validated). Bringing decision scientists together with computer scientists would be highly productive for both fields.

# 4 Recommendations for action

The workshop generated calls for action in four critical to the dealing with emerging challenges in the quantification, communication, and interpretation of uncertainty in simulation and data science:

## 4.1 Uncertainty quantification in large-scale systems

*Transition research in uncertainty quantification of computational systems from the analysis of components to the analysis of large-scale systems of interacting components.*

Existing methodology and current research efforts have focused primarily on individual analyses of small to moderate size. In order to meet the challenge of uncertainty quantification for large-scale systems, a transition is needed to address both very large analyses and large-scale systems of interacting components. A key initial step in addressing uncertainty in large systems will be to develop a representation of uncertainty that can be used to quantify and communicate uncertainty across a broad set of computational problems in large-scale settings in a manner that can be carried through multiple computational processing steps. A framework for addressing uncertainty is needed that can capture at least some level of the distributional behavior of interest beyond that normally captured by simple summary statistics to convey both typical behavior as well as extreme/rare behavior and abrupt changes or tipping points. The framework must facilitate meaningful capture and transfer of information that can be processed efficiently and carried throughout all subsequent computations. In some cases, shared infrastructure may be required to capture all relevant pieces of a large-scale analysis or process, and it will be critical that data be available in a usable form and a timely manner. Both input data and calculated results will require validation processes to ensure data quality and integrity. In predictive settings, discrepancies between actual data values and predicted values will need to be characterized and monitored to provide feedback on unexpected or unexplained results that may require additional investigation.

## 4.2 Uncertainty quantification in data science

*Initiate a major new research and academic initiative in uncertainty quantification for data science.*

There is a clear need to foster work on establishing generalizable methods, standards, and tools for uncertainty quantification for data science, patterned after work that has been done in uncertainty quantification for simulation science. This will require significant progress in four areas: (1) Critical to appropriate handing of uncertainty in data science is the development of methods for measuring and quantifying the various sources of uncertainty in large data sets. These methods must be able to represent and analyze uncertainty due to measurement error, missing data, erroneous data, biased data collection, errors incurred via data integration, and more. As a result, progress in this area will require close cooperation between data scientists and those collecting original source material. (2) New techniques need to be developed for scaling uncertainty quantification in data mining and machine learning to very large data sets. For example, Bayesian methods are far harder to scale than models using simple, linear regression. (3) Principled methods for composing uncertainty across multiple tools used in a processing pipeline are needed. Each step of the processing pipeline introduces additional uncertainties to the final outputs, and these uncertainties need to be correctly propagated so that their cumulative effect is known. (4) Many universities now offer degrees or concentrations in data science. Topics related to the quantification of uncertainty, such as Bayesian statistics, basic models of errors and uncertainty, theory for quantification of the uncertainty induced by approximation techniques such as sub-sampling need to be included in these courses of study. Funding agencies should encourage adequate training through a combination of direct curriculum support and the inclusion of meaningful educational outreach in supported research efforts.

## 4.3 Software support for uncertainty computation

*Create programming systems and tools that facilitate the developed of software involving the representation and analysis of uncertainty.*

The trend will be for greater use of large-scale computational and data-based analyses to aid in decision making. This increased use of computationally-aided decision making will result an a significantly greater demand for programming systems that accommodate uncertainty and error in systematic ways. Substantial research on programming language foundations and practical programming systems is needed to help data scientists develop efficient and correct applications. Thus, we anticipate programming systems and tools that provide frameworks for incorporating uncertainty into existing algorithms, as codes get revised, refactored, updated, etc. The challenges to the development of such systems are both technical and conventional. Technical challenges are the representation of uncertainty in a variety of applications and circumstances as well as the efficient propagation, composition, and estimation of probability distributions and associated parameters. Conventional challenges are exacerbated, for example, when designing software interfaces and developing tools that give developers access to appropriate representations of uncertainty at various levels in the computational process.

### 4.4 Effective communication of uncertainty to stakeholders

*Launch a major new research initiative in communicating uncertainty about large-scale systems to stakeholders.*

To deliver better decisions, all stages of the data-to-decision pipeline must be tightly integrated with improved ways of quantifying uncertainty in both simulation and data science. The importance of accommodating human behavior in models of uncertainty in complex systems needs to be recognized. Uncertainty quantification must be informed by the needs of the users who consume the information and there are at least three critical stakeholder categories: (1) scientists and engineers developing computational models and using such models in their research;

(2) policy makers and others charged with making evidence-based decision, and (3) the general public. Differing levels of expertise, experience, and goal will require differing ways of modeling and communicating uncertainty to each group. To implement this new initiative in communicating uncertainty, funding agencies should provide resources that make genuinely collaborative research attractive. To achieve this goal, we recommend both: (a) creating a fund for simulation and data scientists to add decision scientists to their research groups for the extended periods of time needed to create common language, and (b) creating a center dedicated to integrating decision, simulation, and data science, at a single institution where conversations will happen naturally and frequently.

## REFERENCES

[1] A380 Wing Modifications 2013. Certification of A380 wing rib feet modifications. *Doric Aviation.*

[2] Andre, A. D. and Wickens, C. D. 1995. When users want what's not best for them. *Ergonomics in Design: The Quarterly of Human Factors Applications 3,* 4, 10–14.

[3] Bailey, R. W. 1993. Performance vs. preference. In *Proceedings of the Human Factors and Ergonomics Society 37th Annual Meeting.* 282–286.

[4] Belia, S., Fidler, F., Williams, J., and Cumming, G. 2005. Researchers misunderstand confidence intervals and standard error bars. *Psychological Methods 10,* 4, 389–396.

[5] Bendler, J., Wagner, S., Brandt, T., and Neumann, D. 2014. Taming uncertainty in big data. *Business & Information Systems Engineering 6,* 5, 279–288.

[6] Bonneau, G.-P., Hege, H.-C., Johnson, C. R., Oliveira, M. M., Potter, K., Rheingans, P., and Schultz, T. 2014. Overview and state-of-the-art of uncertainty visualization. In *Scientific Visualization, Mathematics and Visualization,* C. D. Hansen, M. Chen, C. R. Johnson, A. E. Kaufman, and H. Hagen, Eds. 3–27.

[7] Bornholt, J., Mytkowicz, T., and McKinley, K. S. 2014. Uncertain(*T*): A first-order type for uncertain data. In *Proceedings of the 19th International Conference on Architectural Support for Programming Languages and Operating Systems.* ASPLOS '14. ACM, New York, NY, USA, 51–66.

[8] Chandola, V., Banerjee, A., and Kumar, V. 2009. Anomaly detection: A survey. *ACM Comput. Surv. 41,* 3 (July), 15:1–15:58.

[9] Chatfield, C. 1995. Model uncertainty, data mining and statistical inference. *J. of the Royal Statistical Society 158,* 3, 419–466.

[10] Correll, M. and Gleicher, M. 2014. Error bars considered harmful: Exploring alternate encodings for mean and error. *IEEE Transactions on Visualization and Computer Graphics 20,* 12.

[11] Dalvi, N. and Suciu, D. 2007. Efficient query evaluation on probabilistic databases. *The VLDB Journal 16,* 4 (Oct.), 523–544.

[12] de Bruin, W. B. and Bostrom, A. 2013. Assessing what to address in science communication. *Proceedings of the National Academy of Sciences 110,* Supplement 3, 14062–14068.

[13] de Bruin, W. B., Stone, E. R., Gibson, J. M., Fischbeck, P. S., and Shoraka, M. B. 2013. The effect of communication design and recipients' numeracy on responses to UXO risk. *Journal of Risk Research 16,* 8, 981–1004.

[14] Effect of Financial Crisis 2013. The effects of the financial crisis are still being felt, five years on. *The Economist* http://www.economist.com/news/schoolsbrief/21584534-effects-financial-crisis-are-still-being-felt-five-years-article.

[15] Finger, R. and Bisantz, A. M. 2002. Utilizing graphical formats to convey uncertainty in a decision-making task. *Theoretical Issues in Ergonomics Science 3,* 1, 1–25.

[16] Fischhoff, B., Brewer, N. T., and Downs, J. S. 2012. *Communicating Risks and Benefits: An Evidence Based User's Guide.* U.S. Department of Health and Human Services, Food and Drug Administration.

[17] Fischhoff, B. and Davis, A. L. 2014. Communicating scientific uncertainty. *Proceedings of the National Academy of Sciences 111,* Supplement 4, 13664–13671.

[18] Fischhoff, B. and Kadvany, J. 2011. *Risk: A very short introduction.* Oxford University Press.

[19] Fischhoff, B. and Scheufele, D. A. 2013. The science of science communication. *Proceedings of the National Academy of Sciences 110,* 14031–14032.

[20] Fischhoff, B. and Scheufele, D. A. 2014. The science of science communication II. *Proceedings of the National Academy of Sciences 111,* 13583–13584.

[21] Galesic, M., Garcia-Retamero, R., and Gigerenzer, G. 2009. Using icon arrays to communicate medical risks: overcoming low numeracy. *Health Psychology 28*, 2, 210–216.

[22] Garcia-Retamero, R. and Cokely, E. T. 2013. Communicating health risks with visual aids. *Current Directions in Psychological Science 22*, 5, 392–399.

[23] Giry, M. 1982. A categorical approach to probability theory. In *Categorical Aspects of Topology and Analysis,* B. Banaschewski, Ed. Lecture Notes in Mathematics, vol. 915. Springer Berlin Heidelberg, 68–85.

[24] Hammer, B. and Villmann, T. 2007. How to process uncertainty in machine learning. M. Verleysen, Ed. Proc. Of European Symposium on Artificial Neural Networks (ESANN'2007). d-side publications, 79–90.

[25] Han, J. and Orshansky, M. 2013. Approximate computing: An emerging paradigm for energy-efficient design. In 18th *IEEE European Test Symposium (ETS).* 1–6.

[26] Hegarty, M. 2011. The cognitive science of visual-spatial displays: Implications for design. *Topics in Cognitive Science 3*, 3, 446–474.

[26] Hegarty, M., Smallman, S., H., and Stull, A. T. 2012. Choosing and using geospatial displays: Effects of design on performance and metacognition. *Journal of Experimental Psychology:* Applied 18, 1, 1–17.

[28] Howard, R. A. and Matheson, J. E. 2005. Influence diagrams. *Decision Analysis 2*, 3, 127–143.

[29] III, H. D. and Marcu, D. 2011. Domain adaptation for statistical classifiers. *CoRR abs/1109.6341.*

[30] Joslyn, S. and LeClerc, J. 2013. Decisions with uncertainty: The glass half full. *Current Directions in Psychological Science 22*, 4, 308–315.

[31] Kahneman, D. 2011. *Thinking, fast and slow.* Farrar, Giroux and Strauss, New York.

[32] Katrina Forecasters 2005. Katrian forecasters were remarkably accurate. NBC news.

[33] Kinkeldey, C., MacEachren, A. M., and Schiewe, J. 2014. How to assess visual communication of uncertainty? A systematic review of geospatial uncertainty visualisation user studies. *The Cartographic Journal 51*, 4, 372–386.

[34] Loewenstein, G. F., Weber, E. U., Hsee, C. K., and Welch, N. 2001. Risk as feelings. *Psychological Bulletin 127*, 2, 267–286.

[35] Lord Oxburgh. 2010. Scientific assessment panel report. University of East Anglia.

[36] MacEachren, A. M., Robinson, A., Hopper, S., Gardner, S., Murray, R., Gahegan, M., and Hetzler, E. 2005. Visualizing geospatial information uncertainty: What we know and what we need to know. *Cartography and Geographic Information Science 32*, 3, 139–160.

[37] Morgan, M. G. and Henrion, M. 1990. *Uncertainty: A Guide to Dealing with Uncertainty in Quantitative Risk and Policy Analysis.* Cambridge University Press, New York.

[38] National Hurricane Center. 2009. NHC track and intensity models. http://www.nhc.noaa.gov/modelsummary.shtml.

[39] National Research Council. *Intelligence Analysis for Tomorrow: Advances from the Behavioral and Social Sciences.* The National Academies Press, Washington, DC.

[40] National Research Council. 2012. *Assessing the Reliability of Complex Models: Mathematical and Statistical Foundations of Verification, Validation, and Uncertainty Quantification.* The National Academies Press, Washington, DC.

[41] O'Hagan, A., Buck, C. E., Daneshkhah, A., Eiser, J. R., Garthwaite, P. H., Jenkinson, D. J., Oakley, TJ. E., and Rakow, T. 2006. *Uncertain judgements: Eliciting experts' probabilities.* John Wiley & Sons, Chichester.

[42] Okan, Y., Garcia-Retamero, R., Cokely, E. T., and Maldonado, A. 2012. Individual differences in graph literacy: Overcoming denominator neglect in risk comprehension. *Journal of Behavioral Decision Making 25*, 4, 390–401.

[43] Pang, A. T., Wittenbrink, C. M., and Lodha, S. K. 1997. Approaches to uncertainty visualization. *The Visual Computer 13*, 8, 370–390.

[44] Potter, K., Rosen, P., and Johnson, C. 2012. From quantification to visualization: A taxonomy of uncertainty visualization approaches. In *Uncertainty Quantification in Scientific Computing*, A. Dienstfrey and R. Boisvert, Eds. IFIP Advances in Information and Communication Technology, vol. 377. Springer Berlin Heidelberg, 226–249.

[45] Ramsey, N. and Pfeffer, A. 2002. Stochastic lambda calculus and monads of probability distributions. In *Proceedings of the 29th ACM SIGPLAN-SIGACT Symposium on Principles of Programming Languages*. POPL '02. ACM, New York, NY, USA, 154–165.

[46] Reyna, V. F. 2004. How people make decisions that involve risk: A dual-processes approach. *Current Directions in Psychological Science 13*, 2, 60–66.

[47] Reyna, V. F., Nelson, W. L., Han, P. K., and Dieckmann, N. F. 2009. How numeracy influences risk comprehension and medical decision making. *Psychological Bulletin 135*, 6, 943–973.

[48] Smallman, H. S. and St. John, M. 2005. Naïve realism: Misplaced faith in realistic displays. *Ergonomics in Design 13*, 14–19.

[49] Spiegelhalter, D., Pearson, M., and Short, I. 2011. Visualizing uncertainty about the future. *Science 333*, 6048, 1393–1400.

[50] Stone, E. R., Yates, J. F., and Parker, A. M. 1997. Effects of numerical and graphical displays on professed risk-taking behavior. *Journal of Experimental Psychology: Applied 3*, 4, 243–256.

[51] The Tokyo Electric Power Company. Fukushima nuclear accident investigation report (interim report—supplementary volume).

[51] Todd, P. M. and Gigerenzer, G. 2007. Environments that make us smart: Ecological rationality. *Current Directions in Psychological Science 16*, 3, 167–171.

[53] Topchy, A., Jain, A., and Punch, W. 2005. Clustering ensembles: models of consensus and weak partitions. *Pattern Analysis and Machine Intelligence, IEEE Transactions on 27*, 12 (Dec), 1866–1881.

[54] Tversky, A. and Kahneman, D. 1974. Judgment under uncertainty: Heuristics and biases. *science 185*, 4157, 1124–1131.

[55] Vajda, S. 2014. *Probabilistic Programming*. Academic Press.

[56] Weber, E. U. and Stern, P. C. 2011. Public understanding of climate change in the United States. *American Psychologist 66*, 4, 315–328.

[57] Wingate, D., Stuhlmüller, A., and Goodman, N. D. 2011. Lightweight implementations of probabilistic programming languages via transformational compilation. In P*roc. 14th International Conference on Artificial Intelligence and Statistics*.

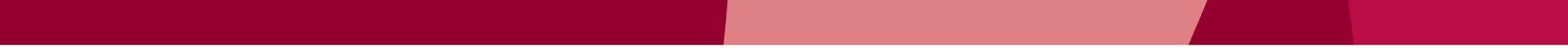
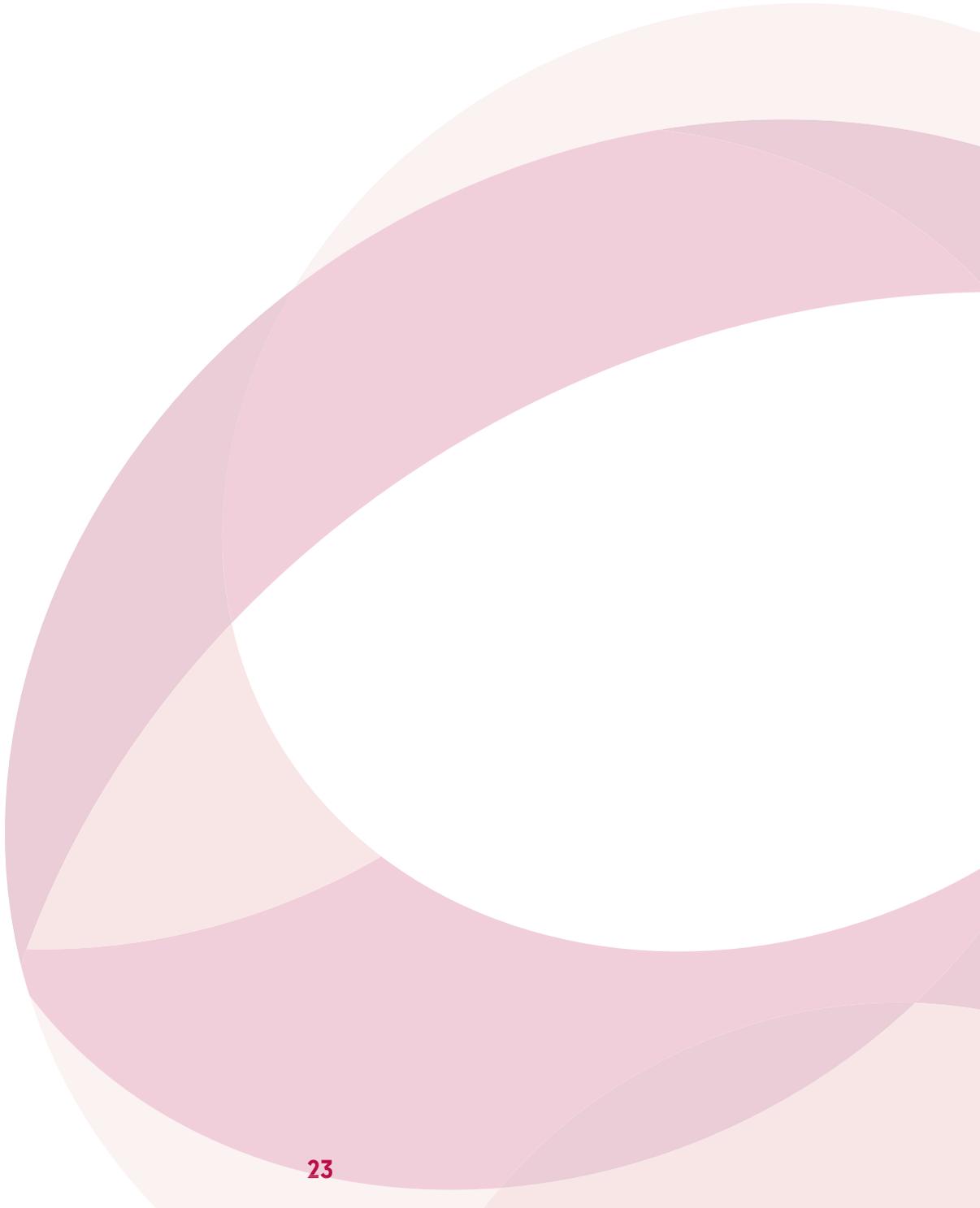

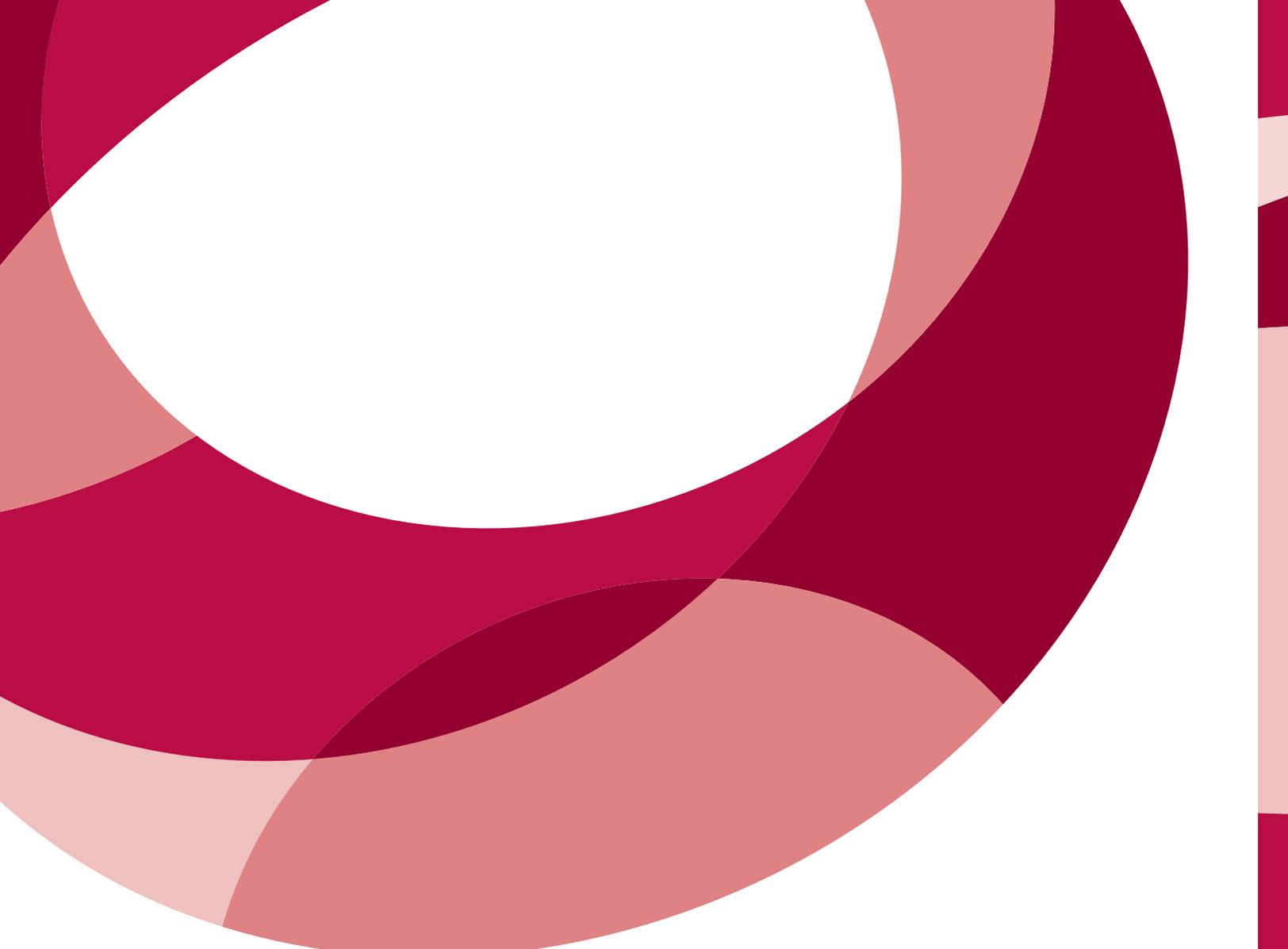

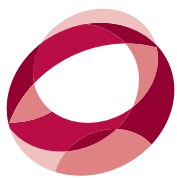 CCC

**Computing Community Consortium**
Catalyst

1828 L Street, NW, Suite 800
Washington, DC 20036
P: 202 234 2111 F: 202 667 1066
www.cra.org cccinfo@cra.org